\documentclass[12pt,preprint]{aastex}

\slugcomment{submitted to Ap.J.}

\shorttitle{ Metallic fingers in exoplanets host stars
}
\shortauthors{Sylvie Vauclair}

\begin{document}


\title{Metallic fingers and metallicity excess in exoplanets host stars : the accretion
hypothesis revisited }


\author{Sylvie Vauclair } 
\affil{Laboratoire d'Astrophysique, Observatoire Midi-Pyr\'en\'ees,
14 avenue Edouard Belin, 31400 Toulouse, France}

%% are identified by the \altaffilmark after each name.  Specify
%%alternate

%% Mark off your abstract in the ``abstract'' environment. In the
%%manuscript
%% title and affiliation information. No date will appear since the
%%author

\begin{abstract}

While the metallicity excess observed in the central stars of planetary
systems is
confirmed by all recent observations, the reason for this excess is still a subject of
debate : 
is it primordial, is it the result of accretion or both? The basic
argument against
an accretion origin is related to the mass of the outer convective zones
which
varies by more than one order of magnitude among the considered stars while the
observed overabundances 
of metals are similar. We show here that in previous discussions a
fundamental
process was forgotten : thermohaline convection induced by the inverse
$\displaystyle \mu $-gradient. ``Metallic fingers" may be created which dilute the accreted matter inside the star. Introducing this effect may reconcile 
the overabundances expected in case of accretion with the observations in stars of different masses. 
\end{abstract}

%% Keywords should appear after the \end{abstract} command. The
%%uncommented



\section{Introduction}

The metallicity distribution of the central stars of planetary systems
compared with other stars of the same spectral types clearly show a
metallicity excess of a factor two on the average, while the individual
[Fe/H] values lie between -0.3 and +0.4 (Santos et al 2003 and
references
therein). Although this excess is generally discussed in terms of iron
abundances only (for most observers [Fe/H] is a synonymous of
``metallicity"), 
spectroscopic observations show that the
abundances of other heavy elements beginning with carbon are also
enhanced
(Santos et al 2001)

The proposed possible explanations for this behavior are of two kinds:

1) accretion hypothesis : this assumes that the star had normal
abundances
at its formation, but that it accreted metal-rich matter during the
phases
of planetary formation (Gonzalez 1998)

2) primordial hypothesis : this assumes that the stellar gas out of
which 
the planetary system formed was already metal-enhanced. A corollary in
this
case would be that such a metal enrichment is necessary for 
planets
to form around stars

Other explanations have been proposed which mix the two previous
scenarios.

The strongest argument invoked against the accretion hypothesis is
related 
to the mass of the outer convective zone which varies by more than one
order of 
magnitude for stars between 1.0 and 1.4 M$_{\odot}$, while the
metallicity 
excess is the same. On the other hand the primordial hypothesis has to
deal
with the fact that some of the host stars have a sub-solar metallicity,
which is an argument against the assumption that metallicity excess is
needed
for planetary formation. 

Diffusion of accreted metals has been discussed by several authors
(Pinsonneault et al 2001, Murray and Chaboyer 2002).
They show that 
gravitational settling is not rapid enough to produce a decrease of the
metal
abundances and lead to the observed values. They also discuss
rotational 
and other mixing processes in relation with the observed lithium
abundances.
As lithium is detroyed by thermonuclear reactions at a relatively small
temperature ($3.10^6$K) inside stars, it is often used as a test
for mixing processes. However such a comparison between the metal
diffusion
and the lithium depletion is highly model dependent as it assumes a
specific
variation of the mixing effect with depth. 

Here we show that the argument against accretion does not hold as soon
as 
we take into account the convective instability induced by inverse 
$\displaystyle \mu $-gradients, or thermohaline convection (also called
``double-diffusive convection"). If freshly accreted metals accumulate
in a small convective zone on top of radiative layers, they will not
stay there but will be diluted downwards in ``metallic fingers" similar to the ``salt fingers" observed in the ocean, so that their overabundance will
rapidly
decrease with time. 

We discuss how this process should take place, with the conclusion that the
 final metallic abundances should not depend on the depth of the standard 
convective zone as usually assumed. However the exact amount of metals which may 
remain in the stellar outer layers depends on parameters like the size and depth 
of the metallic fingers, which cannot be precisely constrained in the framework 
of our present knowledge. We only show that a final overabundance by a factor 
two as observed can be obtained with plausible values of the unknown parameters. 
Two ways can be explored to go further in this study : numerical simulations of 
metallic fingers and asteroseismology of overmetallic stars.

\section{Thermohaline convection}
\subsection{The salted water case}

Thermohaline convection is a wellknown process in oceanography : warm
salted
layers on the top of cool unsalted ones rapidly diffuse downwards even
in the
presence of stabilizing temperature gradients. When a blob is displaced
downwards,
it wants to go further down due to its overweight compared to the
surroundings, but
at the same time the fact that it is hotter contradicts this tendancy.
When the salt
gradient is large compared to the thermal gradient, salted water
normally mixes down until
the two effects compensate. Then thermohaline convection begins. 
While the medium is
marginally stable, salted blobs fall down like fingers while unsalted 
matter goes up around. This process is commonly known as ``salt fingers"
(Stern 1960,
Kato 1966, Veronis 1965, Turner 1973, Turner and Veronis 2000, 
Gargett and Ruddick 2003). The
reason why the medium is still
unstable is due to the different diffusivities of heat and salt. A warm
salted blob falling down
in cool fresh water sees its temperature decrease before the salt has
time to diffuse
out : the blob goes on falling due to its weight until it mixes with the
surroundings.

The salt finger instability can occur with any two components which have
different diffusivities
if there is an unstable gradient of the slower diffusive component and a
stable gradient
of the faster diffusive component. For this reason it is also referred
to as 
``double diffusive convection". In oceanography, the fastest diffusive
component is
conventionally referred to a $T$ while the slowest one is $S$ even when
the components are
different. Indeed the effect is generally studied in the laboratory
using water with a mixture 
of salt and sugar. In this case sugar is the slowest component as it
diffuses about three times slower 
than salt. The results can well be visualized, pictured and studied
(e.g. Wells 2001). The fingers have also been studied by
2D and 3D numerical simulation (e.g. Piacsek and Toomre 1980, Shen and Veronis 1997,
 Yoshida and Nagashima 2003)

The condition for the salt fingers to develop is related to the
density variations induced by temperature and salinity perturbations.
Two important characteristic numbers are defined :

$\bullet$ the density anomaly ratio 
\begin{equation}
R_{\rho} = \alpha \nabla T / \beta \nabla S
\end{equation}
 where 
$\alpha = - (\frac{1}{\rho } \frac{\partial \rho}{\partial T})_{S,P}$ 
and $\beta = (\frac{1}{\rho } \frac{\partial \rho}{\partial S})_{T,P}$ 
while
$\nabla T$ and $\nabla S$ are the average temperature and salinity gradients in the
considered zone

$\bullet$ the so-called ``Lewis number" 
\begin{equation}
\tau = \kappa_S/\kappa_T = \tau_T/\tau_S
\end{equation}
where $\kappa_S$ and $\kappa_T$ are the saline and thermal diffusivities
while $\tau_S$ and $\tau_T$ are the saline and thermal diffusion time scales.

The density gradient is unstable and overturns into dynamical convection
for $R_{\rho} < 1$ while the salt fingers grow for $R_{\rho} \geq 1$. On the other hand 
they cannot form
if $R_{\rho}$ is larger than the
ratio of
the thermal to saline diffusivities $ \tau^{-1} $ as in this case the salinity
difference between the blobs
and the surroundings is not large enough to overcome buoyancy (Huppert and Manins 1973, Gough and Toomre 1982,
Kunze 2003).

Salt fingers can grow if the following condition is satisfied : 
\begin{equation}
1 \leq  R_{\rho} \leq \tau^{-1} 
\end{equation}
In the ocean, $\tau$ is typically 0.01 while it is 1/3 for a salt-sugar
mixture. We will see
below that in solar-type stars where $T$ is the temperature while $S$ is
the mean molecular weight
this ratio is about $10^{-10}$ if $\kappa_S$ is the molecular (or ``microscopic")
diffusion coefficient but it can go up by many
orders
of magnitude
when the shear flow instabilities which induce
mixing between the edges of the fingers and 
the surroundings are taken into account.

\subsection{The stellar case}

Thermohaline convection may occur in stellar radiative zones when a
layer with a larger 
mean molecular 
weight sits on top of layers with smaller ones (Kato 1966, Spiegel 1969,
Ulrich 1972,
Kippenhahn et al 1980). In this case $\nabla_{\mu}$ = dln$\mu$/dln$P$
plays
the role of the
salinity gradient while the difference $\nabla_{ad} - \nabla$
(where 
$\nabla_{ad}$ and $\nabla$ are the usual adiabatic and local (radiative)
gradients 
dln$T$/dln$P$) plays the role
of the temperature gradient. When $\nabla_{ad}$ is smaller than
$\nabla_{rad}$, the temperature gradient
is unstable against convection (Schwarszchild criterion) which
corresponds to warm water below
cool water in oceanography. In the opposite case the temperature
gradient is stable but the medium
can become convectively unstable if (Ledoux criterion):
\begin{equation}
\nabla_{crit} = \frac{\phi}{\delta}\nabla_{\mu} + \nabla_{ad} - \nabla < 0  
\end{equation}
where $\phi=(\partial$ ln $\rho/\partial$ ln $\mu)$ and $\delta=(\partial$ ln $\rho/\partial$ ln $T)$
When this situation occurs, convection first takes place on a
dynamical time scale and
the $\mu$ enriched matter mixes down with the surroundings until
$\nabla_{crit}$ vanishes.
Then marginal stability is achieved and thermohaline convection may
begin as a ``secular process",
namely on a thermal time scale (short compared to the stellar
lifetime!).

Such an effect has previously
been studied for stars with a helium-rich accreted layer (Kippenhahn et
al 1980).
It was also invoked for helium-rich stars in which helium is
supposed to
accumulate due to diffusion in a stellar wind (as proposed by Vauclair
1975) and for roAp stars 
in case some helium accumulation occurs (Vauclair et al 1991). Similar
computations 
have been done in the case of accretion of matter on white dwarfs, 
in relation with novae explosions (e.g. Marks and Sarna 1998).

As we will show below, if hydrogen poor matter is accreted on the top of
a main-sequence type star with normal abundances, it creates an inverse
$\displaystyle \mu $-gradient which may lead to thermohaline convection.
Comparing the stellar case with the water case, we can guess that metallic 
fingers will form if the following condition is verified :
\begin{equation}
1 \leq |\frac{\delta (\nabla_{ad} - \nabla)}{\phi (\nabla_{\mu})}| 
\leq \tau^{-1} 
\end{equation}
with $\tau = D_{\mu} / D_T$  = $ \tau_T / \tau_{\mu}$ where $D_T$ 
and $D_{\mu}$ are the thermal and molecular diffusion
coefficients while 
${\tau_T}$ and $\tau_{\mu}$ are the corresponding time scales.
This condition is similar to
condition (3) for the stellar case. 
In the following, we will neglect the deviations from perfect
gas law, so that $\phi = \delta = 1$. 

In the next section we show computations of the time scales and orders of
magnitude
of this process. 

\section{ The fate of accreted metals in solar type stars}

We study the case of main-sequence solar type stars
which would have
accreted hydrogen-poor material at the beginning of their lifetime. We
assume, for simplicity,
that the accretion occurred in a very short time scale compared to
stellar evolution. We then
study the fate of the accreted metals and chose the examples of 1.1 
M$_{\odot}$ and 1.3 M$_{\odot}$ stars.

As the chemical composition of the accreted matter is not known, we
assume that all
elements are accreted with solar relative abundances except hydrogen and
helium which are
assumed completely absent. We will see that our general conclusion would
not be changed if
the relative abundances were modified : the time scales would only be
slightly different.

After accretion, the metal dilution occurs in two phases. 
Phase one : rapid convection takes place in a dynamical time scale
until it reaches the marginal equilibrium obtained when $\nabla_{crit}$
vanishes (Equation 4 ). 
Phase two : 
thermohaline mixing begins and dilutes the metal excess until 
condition (5)
is no longer satisfied.

In the following, we first discuss the $\displaystyle\mu$-gradients
induced by the original 
metal excess in the 
convective zone. We compute the situation at the end of phase one,
according to the value
of the accreted mass. We discuss the case of the same accretion
occurring for the two 
different stellar masses. Finally we study the time scales of
thermohaline mixing and the
remaining overabundances in the convective zone.

\subsection{ Dynamical convection and marginal equilibrium}

The main parameters of the two stellar models we have used as an example are given in table 1.
They correspond to 1.1 M$_{\odot}$ and 1.3 M$_{\odot}$, both at an age
of about 1.5 Gyr.
We can see that the convective zone is 7 times more massive for the 1.1
M$_{\odot}$ model
than for the 1.3 M$_{\odot}$ model, which does not change sensitively
during 
main sequence evolution. We thus expect that metal accretion leads to an
original
overabundance 7 times larger in the 1.3 M$_{\odot}$ model than in the 1.1
M$_{\odot}$
model.

For completely ionised hydrogen and helium, the mean molecular weight
may be 
obtained with the following expression :
\begin{equation}
\mu = \frac{1 + 4 (\frac{He}{H}) + A (\frac{M}{H})}{2 + 3 (\frac{He}{H})
+ x (\frac{M}{H})} 
\end{equation}
Here A represents an average mass of metals, $(\frac{M}{H})$ a relative
abundance with respect
to hydrogen, and $x$ an averaged number of particules (ion and electrons)
associated with
the metals. 
In the following we take $(\frac{He}{H}) = 0.1$
and $(\frac{M}{H})_0 = 1.4.10^{-3}$ for the original abundances of helium 
and metals (Grevesse and Sauval 1998). We treat the metals as an average 
element to which we attribute the mass $A = 20$. We assume hydrogen and helium 
completely ionized and we neglect the term $x(\frac{M}{H})$.
In these conditions we obtain $\mu_0 = 0.6$ for the value of the mean 
molecular weight before accretion. If metals are added, $\mu$ is modified by :
\begin{equation}
\Delta\mu \simeq 9 \Delta(\frac{M}{H}) 
\end{equation}

Let us now write :
\begin{equation}
 (\frac{M}{H}) =  \alpha (\frac{M}{H})_0  
\end{equation}
and define $\alpha_i$ as the initial value of $\alpha$, obtained just
after
the accretion process (assumed rapid compared to the other time scales).

The variations of $(\frac{M}{H})$ and $\displaystyle \mu$ are then related
by :
\begin{equation}
  \Delta\mu = 9 (\alpha-1) (\frac{M}{H})_0  
\end{equation}

We have computed, as a function of the initial overabundance ratio $\alpha_i$, the
depth at which
metal-enriched material has been diluted when it reaches the marginal
equilibrium
phase, and the actual overabundance ratio $\alpha$ in the convective zone. This has
been obtained
by integrating in all cases the mass of metals diluted by this process.

Suppose, for example, that the metal excess ratio in a 1.1 M$_{\odot}$
star is $\alpha = 2$ after
dilution by dynamical convection. 
This corresponds to an original overabundance factor of 2.05. The same accretion mass would lead
to an original metal excess ratio $\alpha_i = 14.3$ in a 1.3 M$_{\odot}$ star.
Our computations show that such an original metal excess is diluted
in a ``transition zone" down to $r_{TZ}= 7.5.10^{10}$cm, 
leading to an overabundance after dilution $\alpha = 8.5$. 
Figure 1 displays the abundance profiles obtained in this case
in the 1.3 M$_{\odot}$ star.
The 
parameters at the bottom
of the transition zones as we have defined above 
 are shown in table 1 for the two models. 
Figure 2 and 3 display, for the two models, the values 
of $\alpha_i$ and the corresponding $\alpha$ as a function 
of the width of the transition zone.

Thermohaline mixing begins after this dilution process.

\clearpage

\begin{deluxetable}{lrrrrcrrrrr}
\tablewidth{0pt}
\tablecaption{Model data$^a$}
\tablehead{
\colhead{model}           & \colhead{$r$}      &
 \colhead{$r/R_*$}      &
\colhead{$\Delta M_r$}          & \colhead{$T_r$}  &
\colhead{$\rho_r$}          & \colhead{$P_r$}    &
\colhead{$\kappa_r$}  & \colhead{$\tau_{th}$(yr)}  }

\startdata
 $ 1.1$ (CZ) &  $5.2e10$ & $ 0.75$ & $3.7e31$ & $2.0e6$ & $.128$ & $3.5e13$ &
\nodata
& \nodata \\
 $ 1.1$ (TZ) & $4.9e10$ & $ 0.71$ & $4.6e31$ & $2.3e6$ & $.195$ & $6.2e13$ &
$15.8$ &
$4200$ \\
 $ 1.3$ (CZ) & $8.0e10$ & $ 0.84$ &$5.1e30$ & $9.0e5$ & $0.007$ & $1.2e12$ &
\nodata
& \nodata \\
 $ 1.3$ (TZ) & $7.5e10$ & $ 0.79$ & $8.2e30$ & $1.2e6$ & $.013$ & $2.2e12$ &
$12.6$ &
$800$ \\
\enddata
\tablenotetext{a}{the values are given at the bottom of the convective zone
(CZ) and
at the bottom of the ``transition zone" (TZ) defined as the region of
metal dilution after
dynamical convection, for the example discussed in the text (metal excess
$\alpha = 2$ in the 1.1 M$_{\odot}$ model) ; the metal abundances decrease with depth
 inside this region
until it reaches
its normal value at $r_{TZ}$ (cf. Figure 1).}

\end{deluxetable}

\clearpage

\begin{figure}
\plotone{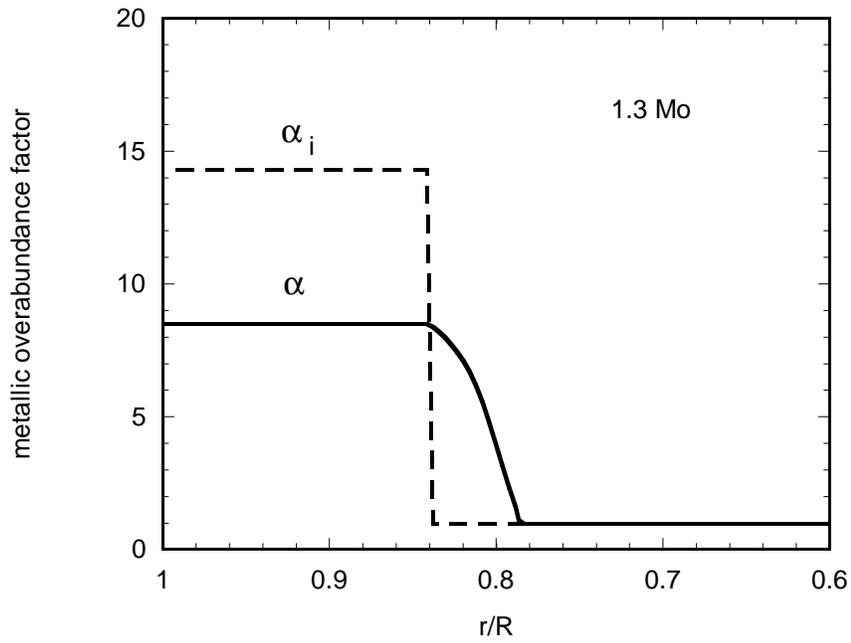}
\caption{Profiles of the metallic overabundance factors in the 
1.3 M$_{\odot}$ model before ($\alpha_i$) and after ($\alpha$) dilution
by dynamical convection for the example discussed in the text, as a function
of the fractional radius inside the star. Thermohaline convection begins after this phase.
 \label{fig1}}
\end{figure}

\clearpage 

\begin{figure}
\plotone{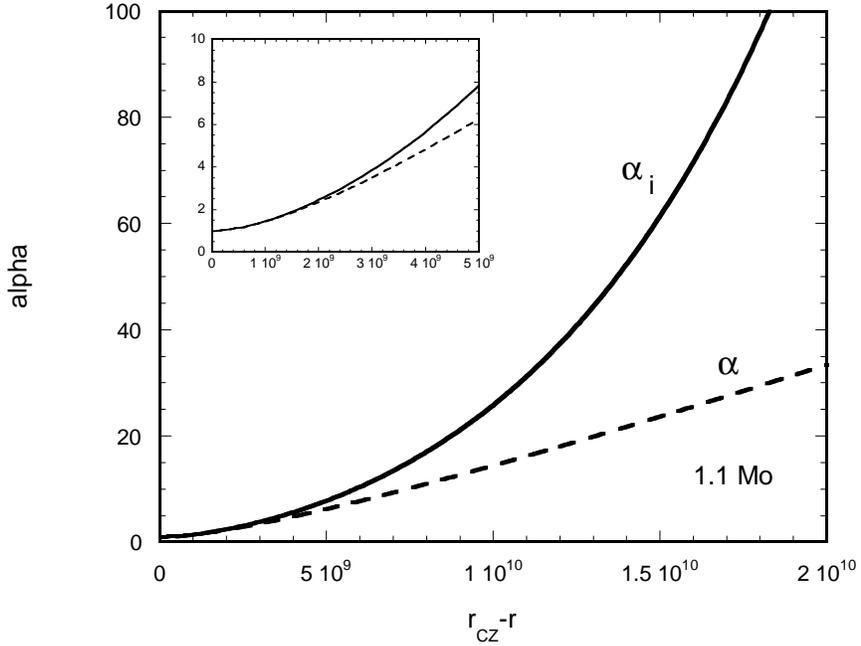}
\caption{This figure shows, for the 1.1 M$_{\odot}$ model, 
the initial metallic excess $\alpha_i = (\frac{M}{H})_i/(\frac{M}{H})_0$
and the metallic excess $\alpha$ obtained after dilution by dynamical convection, 
as a function of the depth 
$(r_{cz}-r)$ at
which
they are diluted. The $\alpha$ curve is obtained with $\nabla_{crit} = 0$ (Equation 4). 
For example, an original overabundance of 30 will be diluted inside a transition region 
of thickness $1.05.10^{10}$cm, below the standard convective zone. 
An overabundance of 15 will remain after this process. 
Then thermohaline convection will go on reducing this value. 
The curves close to the origin are displayed in a zoom.
\label{fig2}}
\end{figure}

\begin{figure}
\plotone{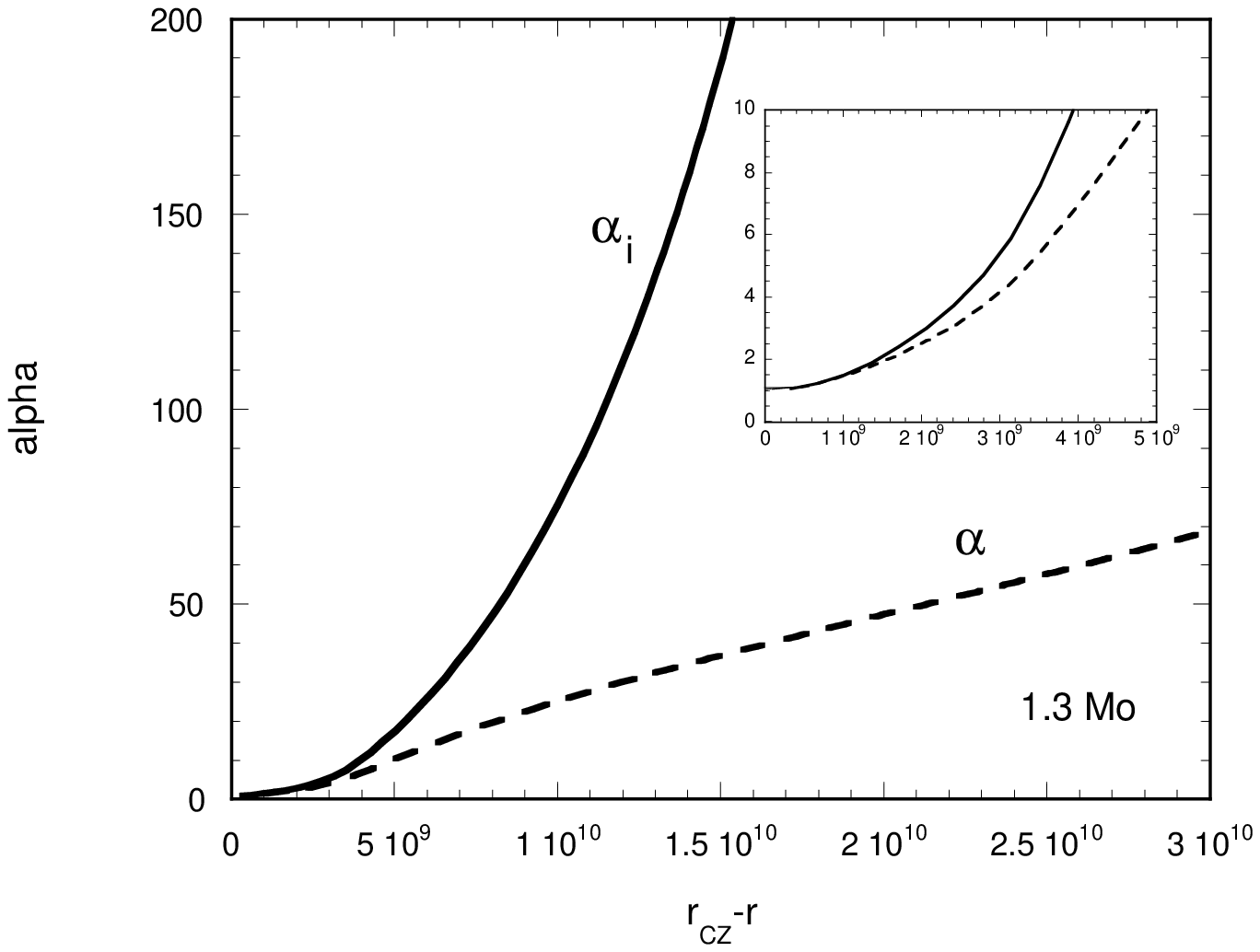}
\caption{Same as Fig 2, for the 1.3 M$_{\odot}$ model 
M$_{\odot}$ 
\label{fig3}}
\end{figure}

\subsection{ Thermohaline mixing, metallic fingers and time scales}

The study of thermohaline mixing in stars is far from trivial. Detailed 
comparisons of numerical simulations and laboratory experiments in the 
water case have recently been published (Gargett and Ruddick 2003) but
the stellar case may be different as mixing then occurs in a compressible
stratified fluid.

In the following we use the formalism 
proposed by 
Kippenhahn et al 1980 (hereafter KRT) who discussed the time scale 
of thermohaline mixing in stars in a simplified way and applied it 
to the case of helium rich
layers standing on the top of
hydrogen rich ones. 

In first approximation they treated thermohaline mixing as a diffusion
theory. The whole picture may be described as follows: 
blobs of metal enriched matter begin to fall down from the convective zone and exchange
heat and heavy elements with their surroundings; chemicals diffuse
more slowly than heat, so that the blobs go on falling down until
they are completely disrupted, thereby creating finger shapes.
The most efficient process for element diffusion out of the blobs
is the shear flow instabilities at the edge of the fingers :
as the falling down matter undergoes friction with matter going up,
turbulence occurs and mixes part of the fingers with their surroundings,
on a horizontal length scale which is a fraction $\epsilon$ 
of the horizontal size $L$ of the blobs. 
The blobs disappear when they have travelled down on a 
distance $W$ long enough for this mixing to disrupt them
completely.

KRT define a diffusion coefficient $D_{th}$ 
as the product of the blobs vertical velocity $v_{\mu}$ 
by their size $L$. 
They evaluate $v_{\mu}$ as :
\begin{equation}
v_{\mu} = \frac{H_p}{(\nabla_{ad} - \nabla)\tau_{KH}^* }(\frac{D\mu}{\mu})
\end{equation}
where $D\mu$ represents the $\mu$ difference between the inside and the outside
of the blobs.
The diffusion coefficient $D_{th}$ is then obtained as :
\begin{equation}
D_{th} = \frac{H_p}{|\nabla_{ad} - \nabla|} \frac{L^2}{\tau_{KH}^*}
|\frac{dln\mu}{dr} |
\end{equation}
where $H_p$ is the pressure scale height and
$\tau_{KH}^*$
the local thermal time scale given by :
\begin{equation}
\tau_{KH}^* = \frac{C_p \kappa\rho^2 L^2}{16 a c T^3} 
\end{equation}
where the parameters have their usual meanings. We can see that $L^2$
vanishes in 
this diffusion coefficient which 
does not depend on the size of the blobs.

In this simple picture, $D_{th}$ is not exactly the local mixing coefficient as
it involves the whole blobs while local mixing involves only the edges of the blobs. 
The local
mixing coefficient becomes correctly represented by $D_{th}$ only
at the bottom of the fingers, where the blobs disappear.
For this reason, KRT define the thermohaline diffusion 
time scale as : $\tau_{th} = W^2/D_{th}$
where $W$ is
the depth of the fingers (it also
represents the size of the transition zone in which $\mu$ varies by $\Delta\mu$,
between 
its two extreme values). This diffusion time scale may be written :
\begin{equation}
\tau_{th} = \frac{\mu}{\Delta\mu} \frac{C_p \kappa\rho^2 W^3}{4 a c
T^3} \frac{\nabla_{ad} - \nabla }{H_p}
\end{equation}
Note that if the mixing scale at the finger edges increases linearly
while the blobs fall down, the coefficient $\epsilon$ is nothing
else than the ratio of the horizontal and vertical length scales
of the fingers, namely $L/W$.

We give in table 1 the thermohaline diffusion 
time scales corresponding to the
example given in the previous section, with a metallic excess of a
factor 2
in 1.1  M$_{\odot}$. 
Computations done with a metal excess of 1.5 instead of 2
 give similar
results for the thermohaline diffusion time scales. They are always very short
compared to the 
stellar lifetime. 
If nothing could stop these metallic fingers from growing, the convective zone would
be
emptied of its metallic matter short after it is accreted.

We must now take into account the second condition for the formation of the fingers
(Equation 5), namely :
\begin{equation}
|\frac{\nabla_{ad} - \nabla}{\nabla_{\mu}}| \leq \ \frac{D_{T}}{D_{\mu}}
\end{equation}
At the end of the dynamical convection phase, the 
$|\frac{\nabla_{ad} - \nabla}{\nabla_{\mu}}|$ ratio is one. 
Then, during thermohaline convection, the local $\mu$-gradient decreases
so that the $|\frac{\nabla_{ad} - \nabla}{\nabla_{\mu}}|$ ratio increases. 
Thermohaline convection should stop when this ratio reaches the
value of the $D_T/D_\mu$ ratio. 
The remaining overabundances in the stellar outer layers
strongly depend on the efficiency of the turbulent mixing at the edge
of the fingers, which is non linear and not really known. 
We can only give here a tentative discussion of this process.

The thermal diffusion coefficient $D_T$ may be evaluated as the square
of a relevant length scale (e.g. the size of the falling blobs) divided
by the local thermal time scale $\tau_{KH}^*$ :
\begin{equation}
D_T = \frac{L^2}{ \tau_{KH}^* } = \frac{16 a c T^3 }{ C_p \kappa\rho^2} 
\end{equation}

Below the convective zone in the 1.1 M$_\odot$ , $D_T$ 
is typically of $3.0.10^{8} $cm$^2.$s$^{-1}$ while
it is $8.10^{9} $cm$^2.$s$^{-1}$ in the 1.3 M$_\odot$ model.
Meanwhile the microscopic element diffusion coefficient is of order 
one cm$^2.$s$^{-1}$ : if no turbulence occured, the metallic fingers 
would extend far down and 
deplete the overmetallic material until an extremely small $\mu$-gradient
would be reached :
in this case 
no metallic excess would be left in
the convective zone.

If we take into account the shear flow instabilities which result 
from the motion of the blobs, $D_{\mu}$ has to be replaced by the
local turbulent diffusion coefficient which is not exactly $D_{th}$
but ${\epsilon}^2 D_{th}$ where ${\epsilon}$ represents the $L/W$ ratio.
Using equation (11) for $D_{th}$, we find from
condition (14) that the fingers disappear when $W$
is of the same order as $L$, which makes sense but is not
helpful to evaluate the amount of heavy elements which
remains in the outer layers at the end of the whole process :
the KRT approximations are too rough to be useful in this respect. 

Another approach consists in estimating the value of $D_{\mu}$
which would be needed to account for the observations.
Our aim is that a metallicity excess of order two remains 
in the convective zone after these processes occurred. From
Equation (9) we find that it corresponds to $\Delta\mu \simeq 0.0126$
or, with $\mu = .6$ as computed in these stellar layers, 
$\Delta\mu / \mu \simeq 0.02$. If we suppose that the mixing region
extends over one pressure scale height (which is by not means proved !) 
we find that the $D_T/D_{\mu}$ ratio should be of order 10, which seems 
reasonable compared to the water case.
Such a value would be obtained with a turbulent diffusion coefficient 
$D_{\mu}$ at the edge of the fingers of 
order $10^7$ to $10^9$ cm$^{2}$.s$^{-1}$. 

More computations are needed for a better understanding of
the shear flow turbulence induced by the fingers. 
Furthermore the growing of
 the metallic fingers may depend on the other processes at work in the star :
rotation-induced turbulence, internal waves, etc. 
(cf. similar problems in salted water : Gargett and Ruddick 2003).

In any case the computations presented here show that the metallic matter 
accreted onto a star does not stay inside the standard convective zone : 
it first turns over due to 
dynamical convection and then goes on diffusing due thermohaline convection. 
The observed overabundances in planetary systems host stars
can be obtained with realistic values of the unknown parameters.

\section{Conclusion and future prospects }

The important conclusion of this paper 
is that the strongest argument 
against accretion as an explanation of the metallic excess observed
in host stars planetary systems host stars has to be revised : if hydrogen
poor matter is accreted in the early phases of stellar evolution, during
planet formation, the metal excess in the convective zone leads to
a destabilizing $\displaystyle \mu $-gradient which induces metal dilution, 
first by dynamical convection, then by thermohaline mixing, 
on a time scale much shorter than the stellar lifetime (typically 1000 yrs). 
The remaining 
overabundances in the convective zones
depend on the physical conditions inside the star. We do not expect, 
in case of accretion, to end up with metallic overabundances 
increasing with decreasing convective mass as obtained in standard models.
Furthermore, the region mixed by the thermohaline process
is localised below the convective zones 
and should not go down to the lithium destruction layers. 

In the present study, we have supposed that the accreted material had a relative 
solar abundance, except for hydrogen and helium which were assumed completely absent. 
Changing these relative abundances would modify the relation between the 
$\mu$ value and the metallicity (Equation 9). 
The general conclusions would still hold : 
the numerical values and time scales would only be changed accordingly.

We did not discuss the amount of matter needed to explain the observed abundances
by accretion and how this accretion could have proceed. 
In the example of section 3.1, a metal excess of 2.05 in 1.1 M$_\odot$ 
star would be obtained with 145 M$_{\oplus}$ of all metals included,
corresponding to about 11 M$_{\oplus}$ 
 of iron, which is a large value. A metal excess of 1.5 would 
be obtained with 5 M$_{\oplus}$ of iron or 66 M$_{\oplus}$ of all metals.
The difficulty of explaining such a large accretion from 
protoplanetary discs represents a second argument against 
the accretion hypothesis, which we do not discuss here :
this is a different subject, out of the scope 
of the present paper. 

Detailed spectroscopic observations and precise abundance 
determinations will be of
great interest to test the accretion processes against the primordial
scenario for exoplanets hosts stars.
Asteroseismic studies of these stars 
 will also help
deriving whether they are overmetallic down to the center or only
in their outer layers and give an important hint for a better understanding
of the formation of planetary systems.

\acknowledgments

I thank the referee N. Murray who gave very useful comments and critics on a first
version of this paper and helped improving the discussion.

%\clearpage

%% The following command ends your manuscript. LaTeX will ignore any
%%text


\begin{thebibliography}{}

\bibitem[]{}Gargett, A., Ruddick, B. 2003, ed., Progress in Oceanography, vol 56,
issues 3-4, pages 381-570, special issues on ``Double diffusion in
Oceanography", 
\bibitem[]{}Gonzalez, G. 1998, A\&A, 334, 221
\bibitem[]{}Gough, D.O., Toomre, J., 1982, J.Fluid Mech, 125, 75
\bibitem[]{}Grevesse, N., Sauval 1998, in ``Solar Composition and its Evolution - from Core to Corona"
ed. C. Fr\''{o}lich, M.C.E. Huber, S.K.Solanki, R. von Steiger, Kluwer Academic Publishers
\bibitem[]{}Huppert, H.E., Manins, P.C. 1973, Deep-Sea Research, 20, 315
\bibitem[]{}Kato, S., 1966, PASJ, 18, 374
\bibitem[]{}Kunze, E. 2003, Progress in Oceanography, 56, 399
\bibitem[]{}Kippenhahn, R., Ruschenplatt, G., Thomas, H.C.
1980, A\&A, 91, 175
\bibitem[]{}Marks, P.B., Sarna, M.J.
1998, \mnras, 301, 699
\bibitem[]{} Murray, N., Chaboyer, B. 2002, \apj, 566, 442
\bibitem[]{} Piacsek, S.A., Toomre, J. 1980, in J.C.J. Nihoul (ed.), Marine turbulence,
(pp193-219), Elsevier oceanography series, 28 
\bibitem[]{} Pinsonneault, M.H., DePoy, D.L., Coffee, M. 2001, \apj,
556, L59
\bibitem[]{}Santos, N.C., Israelian, G., Mayor, M., Rebolo, R., Udry, S.
2003, A\&A, 398, 363
\bibitem[]{}Santos, N.C., Israelian, G., Mayor, M.
2001, A\&A, 373, 1019
\bibitem[]{}Shen, C.Y., Veronis, G. 1997, J. Geophys. Res., 102 (c10), 23131
\bibitem[]{}Spiegel, E. 1969, Comments on Ap and Space Phys ; 1, 57
\bibitem[]{}Stern, M.E., 1960, Tellus, 12, 2
\bibitem[]{}Turner, J.S.1973, Buoyancy effects in fluids, Cambridge University Press
\bibitem[]{}Turner, J.S., Veronis, G. 2000 journal fluid dynamics, 405,
269
\bibitem[]{}Ulrich, R.K., 1972, \apj, 172, 165
\bibitem[]{}Vauclair, S.1975, A\&A, 45, 233
\bibitem[]{}Vauclair, S., Dolez,N., Gough, D.O. 1991, A\&A, 252, 618
\bibitem[]{}Veronis, G.J., 1965, Marine Res.,21, 1 
\bibitem[]{}Wells, M.G., 2001, PhD thesis,``convection, turbulent mixing and
salt fingers", The Australian National University
\bibitem[]{}Yoshida J., Nagashima, H. 2003, Progress in Oceanography, 56, 435




\end{thebibliography}
\end{document}